\documentclass[aps,pre,twocolumn,showpacs,floatfix]{revtex4} 
\usepackage{amsmath,amssymb,graphicx,times,bm,braket}

\newcommand{\be}{\begin{equation}}
\newcommand{\ee}{\end{equation}}
\newcommand{\bea}{\begin{eqnarray}}
\newcommand{\eea}{\end{eqnarray}}
\newcommand{\bi}{\begin{itemize}}
\newcommand{\ei}{\end{itemize}}
\newcommand{\bw}{\begin{widetext}}
\newcommand{\ew}{\end{widetext}}
\newcommand{\kommentar}[1]{}
\newcommand{\pavg}{\Pi(t)}
\newcommand{\tr}{{\rm tr}}

\DeclareMathSymbol{\varTau}{\mathord}{letters}{84}

\begin{document}
 
\title{Environment-assisted quantum transport and trapping in dimers
}
\author{Oliver M{\"u}lken}
\author{Tobias Schmid}
%\email{muelken@physik.uni-freiburg.de}
\affiliation{
Physikalisches Institut, Universit\"at Freiburg,                                                 
Hermann-Herder-Stra{\ss}e 3, 79104 Freiburg, Germany}

\date{\today} 

\begin{abstract} 
We study the dynamics and trapping of excitations for a dimer with an energy off-set $\Delta$ coupled to an external environment.  Using a Lindblad quantum master equation approach, we calculate the survival probability $\Pi(t)$ of the excitation and define different lifetimes $\tau_s$ of the excitation, corresponding to the duration of the decay of $\Pi(t)$ in between two predefined values. We show that it is not possible to always enhance the overall decay to the trap. However, it is possible, even for not too small environmental couplings and for values of $\Delta$ of the order ${\cal O}(1)$, to decrease certain lifetimes $\tau_s$, leading to faster decay of $\Pi(t)$ in these time intervals: There is an optimal environmental coupling, leading to a maximal decay for fixed $\Delta$.
\end{abstract}
\maketitle

\section{Introduction}
Recent years have seen a growing interest of different communities in
coherent energy transfer. For instance, it was noted that in the
light-harvesting process of photosynthesis
quantum mechanical features of the transfer of excitations (excitons)
created by the incoming solar photons might play an important role for the
transfer's efficiency \cite{engel2007,collini2010}. 
Models of the (coherent) exciton dynamics in the light-harvesting complex
of the photosynthetic unit show that the environment does not necessarily
destroy all coherent features -- even at ambient temperature -- but also
can support the coherent transfer of excitations
\cite{cheng2006,mohseni2008,caruso2009,olaya-castro2008,thorwart2009}. 
In particular, Rebentrost {\sl et al.} have given a detailed analysis of
excitonic dynamics in the Fenna-Matthews-Olsen
complex introducing the concept of environment-assisted quantum
transport \cite{rebentrost2009}. They showed that the maximum transport
efficiency is reached for decoherence rates comparable to the difference
between the onsite and the coupling energies. Cao and Silbey showed that
exciton trapping can be optimized by suitable choices of, e.g.,
decoherence and trapping rates \cite{cao2009}. 

Rapid experimental advances in ultra-cold gases allow to control
atoms to a large extent. This offers the possibility to study
coherent transport and the effect of environmental changes (e.g., an
increase in the temperature or of an external field). Rydberg gases turn
out to be very well suited to study the dynamics of excitations
\cite{anderson1998,mourachko1998,westermann2006}. At ultra-low
temperatures the dynamics in an ensemble of atoms can be efficiently
modeled by continuous-time quantum walks \cite{mb2010a}.
By properly adjusting specific single Rydberg atoms, the moving excitation
can be absorbed by these atoms  \cite{mbagrw2007,reetz-lamour2008}. An
analogous process is found in the light-harvesting process, where the
exciton eventually will reach the reaction center, where the exciton's
energy gets absorbed and converted to chemical energy. In both cases, the
decay of the probability $\Pi(t)$ not to get absorbed monitors the excitation's
dynamics. If the process is purely coherent, $\Pi(t)$ shows distinct
quantum-mechanical features such as a power-law decay in certain,
experimentally relevant time-intervals \cite{mbagrw2007,mb2010a}.

Since usually the systems (light-harvesting complexes or Rydberg gases) are not
isolated from their environment, we will model the dynamics
by the Lindblad quantum master equation (LQME) for the reduced
density operator of the system. However, it should be noted that this
approach is only valid in a limited parameter range of the coupling to the
environment \cite{mmsb2010}.
Our model system is a dimer, represented by two coupled two-level systems,
one of which acting as trap.
We note that various systems, e.g., with radial symmetry and a trap in the
center \cite{mbb2006a} or with strong coupling between all nodes and weak coupling
to the trap, can be mapped onto the dimer.

\section{Model}
In general, we consider a small system $S$ (network) of $N$ nodes coupled to an
environment. Each node of the network $S$ represents a single two-level
system. The basis states $|j\rangle$ which are associated with
excitations localized at the nodes $j=1,\dots,N$ span the Hilbert space of
$S$ alone. 
The Hamiltonian of the total system, $\bm
H_{\rm tot}$, comprises three parts: the Hamiltonians $\bm H_S$ for
the network $S$, $\bm H_R$ for the environment (reservoir), and $\bm H_{SR}$
for the coupling between system and environment: $\bm H_{\rm tot} = \bm
H_S + \bm H_R + \bm H_{SR}$.

Within a
phenomenological approach, the Hamiltonian $\bm H_S$, which incorporates
trapping of excitations at the nodes $m \in \mathcal{M}$, $\mathcal{M}
\subset \{1, \dots, N\}$, is given by 
${\bm H}_S \equiv
{\bm H}_0 - i{\bm \Gamma}$,
where $\bm H_0$ is the network Hamiltonian without any trapping and $i{\bm
\Gamma} \equiv i \Gamma \sum_{m} | m \rangle \langle m |$ ($\Gamma>0$) is
the trapping operator, see Ref.~\cite{mbagrw2007} for details. 
%As a result, ${\bm H}_S$ is non-Hermitian and usually has $N$ complex eigenvalues, $E_l = \epsilon_l - i\gamma_l$ ($l=1,\dots,N$) where $\gamma_l>0$, and $N$ right and $N$ left eigenstates, denoted by $|\Phi_l\rangle$ and $\langle\tilde\Phi_l|$, respectively. 

Now, the dynamics of the density operator of the total system is governed by the Liouville-von~Neumann
equation.
Integrating out the environmental degrees of freedom and assuming bilinear couplings between system and environment and the Markov
approximation lead to \cite{Breuer-Petruccione}
\be
\dot{\bm\rho}(t) = -i [\bm H_0 , \bm\rho(t)] - \{\bm\Gamma,\bm\rho(t)\}
+ {\cal D}[\bm\rho(t)],
\label{qme}
\ee
where ${\cal D}[\bm\rho(t)]$ is responsible for all decoherence effects.

Under certain conditons, such as a weak coupling between the system and
the environment, 
%${\cal D}[\bm\rho(t)]$ can also be expressed by operators acting on the Hilbert space of the system $S$. 
Eq.~(\ref{qme}) leads
%can then be brought into the so-called Lindblad form leading 
to the Lindblad quantum
master equation (LQME),
where the Lindblad operators, acting on the Hilbert
space of $S$, mimick the
influence of the environment on the dynamics \cite{Breuer-Petruccione}.
Considering only localized initial conditions
$\bm\rho_k(0)=|k\rangle\langle k|$ and  
Lindblad operators in the form
$\sqrt{\lambda} |j\rangle\langle j|$ which act
only on the diagonal elements of $\bm\rho(t)$, the LQME
reads \cite{mmsb2010}
\bea
\dot{\bm \rho}_k(t) &=& -i \big[ {\bm H_0}, {\bm \rho}_k(t) \big] -
\big\{{\bm\Gamma},{\bm\rho}_k(t) \big\}
\nonumber \\
&&
- 2\lambda
\sum_{j=1}^N \Big(\bm\rho_k(t) - \langle j | \bm\rho_k (t) | j
\rangle\Big) |j\rangle\langle j|,
\label{lvne}
\eea
where the initial condition is labeled at the reduced
density operator by the subscript $k$. The parameter $\lambda$ represents
the ``strength'' of the coupling to the environment.
Now, the transition probabilities $\pi_{k,j}(t)$
from node $j$ to node $k$ follow from the diagonal elements of
$\bm\rho_k(t)$, i.e., $\pi_{k,j}(t) = \langle j| \bm
\rho_k(t) | j\rangle$. 
%When the system is not coupled to the environment ($\lambda = 0$), the probabilities are obtained by direct diagonalization of $\bm H_S$, which yields 
%\begin{equation}
%\pi_{k,j}(t) = \Big|\sum_{l=1}^N \exp(-\gamma_lt) \exp(-i\epsilon_lt)\langle k | \Phi_l \rangle \langle \tilde\Phi_l | j \rangle\Big|^2,
%\end{equation}
%such that in the purely coherent case the negative imaginary parts $\gamma_l$ of $E_l$ determine the temporal decay. 

In general, the mean survival probability is defined as
\be
\pavg \equiv \big\langle \tr[\bm\rho_k(t)] \big\rangle_k =
\frac{1}{N-M}\sum_{k\neq m} \sum_{j=1}^N \langle j| \bm \rho_k(t) |j \rangle,
\label{pavg}
\ee
where $\langle \cdot \rangle_k$ denotes the average over all possible
initial nodes, i.e., all nodes but the trap nodes. Note, that
Eq.~(\ref{pavg}) slightly deviates from the definition used in
\cite{mbagrw2007} for the purely coherent case; there, only
final nodes $j\neq m$ have been considered. Equation~(\ref{pavg}) also
accounts for the fact that an excitation has a probability of going from a
node $m$ to a node $j\neq m$. Thus,
$1-\Pi(t)$ describes the total probability of energy dissipation up to
time $t$.

\section{Results}
\begin{figure}[h]
\centerline{\includegraphics[clip=,width=0.9\columnwidth]{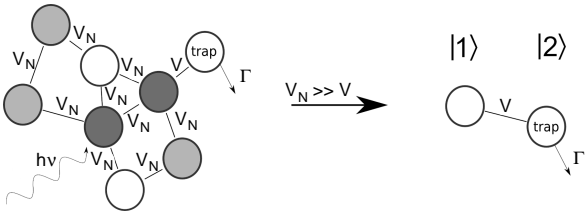}}
\caption{Schematic illustration of the dimer model.}
\label{dimer}
\end{figure}

In order to obtain results which are not blurred by the networks
complexity, we consider a dimer model, which is composed out of two nodes,
one of which acts as the trap. However, it is also
possible to map more complex networks onto an effective dimer: One example is a network where the couplings $V_N$ between the
non-trap nodes are very strong and only a single node is coupled to the
trap with a coupling $V\ll V_N$, see
Fig.~\ref{dimer} or \cite{Amerongen}.
Another example is a ring of nodes which are all (weakly) coupled to a
trap in the center of the ring; if the initial excitation is fully
delocalized over the ring, the effective dynamics can also be modeled by a
dimer: the non-trap node of the dimer is given by a superposition of all
the nodes in the ring and the trap of the dimer is identified with the
trap in the center of the ring.

Now, consider a dimer coupled to an external bath. The Hamiltonian of
the dimer without any coupling to the surroundings can be expressed in
matrix notation by
\be
{\bm H}_S = {\bm H_0} -i \bm \Gamma =
\begin{pmatrix} E_1 & -V\\-V & E_2-i\Gamma \end{pmatrix}
\ee
where $E_1$ and $E_2$ are the on-site energies and $V$ is the coupling
between the two nodes. Note that in the following we will express all
varying parameters in units of $V$.  The eigenvalues of $\bm H$ are
$E_\pm= E_1 \pm V e^{\pm \phi_\Delta}$, 
where $\phi_\Delta = \textrm{arcsinh}[(\Delta - i\Gamma)/2V]$ and
$\Delta=|E_1-E_2|$, and where
we assumed $\Gamma\leq 2V$, such that (for $\Delta=0$) the dimer is not
overdamped \cite{mmsb2010}.
The
bi-orthonormalized eigenstates of ${\bm H}$ are of the form
\begin{equation}
|\Phi_\pm\rangle \equiv \frac{1}{\sqrt{2\cosh\phi_\Delta}} \left( \begin{matrix}
e^{\pm \phi_\Delta/2} \\
\pm e^{\mp \phi_\Delta/2} \end{matrix} \right) 
\end{equation}
and
\begin{equation}
|\tilde\Phi_\pm\rangle \equiv \frac{1}{\sqrt{2\cosh\phi_\Delta^*}} \left(
\begin{matrix} e^{\mp \phi_\Delta^*/2} \\
\pm e^{\pm \phi_\Delta^*/2} \end{matrix} \right),
\end{equation}
where $\phi_\Delta^*$ is the complex conjugate of $\phi_\Delta$.
In the limit $\Delta\to 0$ we recover the results of \cite{mmsb2010}.
Note also, that our model is similar to the one studied by Cao and Silbey
in \cite{cao2009}. There, the authors obtained analytical estimates for
the (quantum) mean first passage time.

\begin{figure}[ht]
\centerline{\includegraphics[clip=,width=0.9\columnwidth]{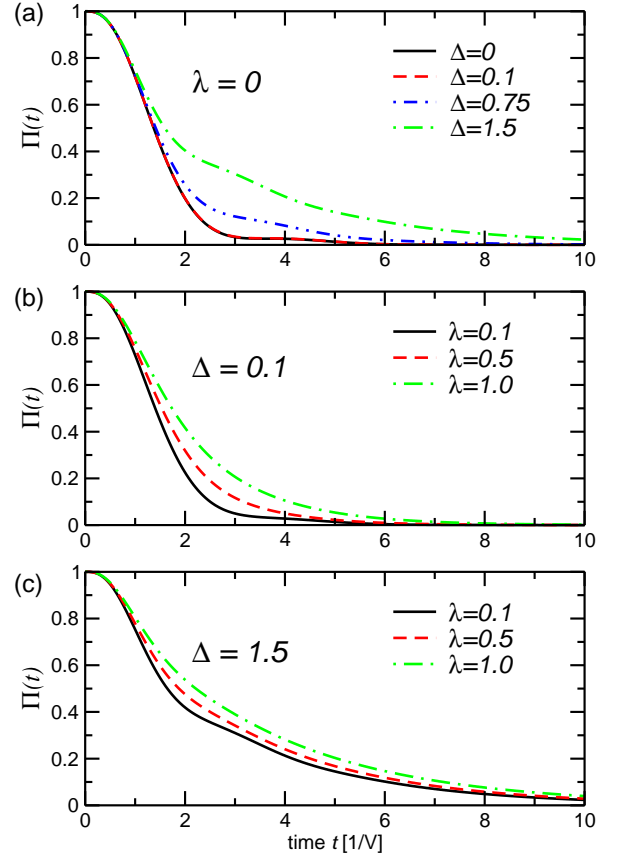}}
\caption{(Color online) Survival probability $\Pi(t)$ for $\Gamma=V=1$:
(a) for $\lambda=0$ and different $\Delta$, see Eq.~(\ref{pil0}). (b)
solution of the LQME for $\Delta=0.1$ and different $\lambda$. (c) same as
(b) but for $\Delta=1.5$.}
\label{surv-probs}
\end{figure}

\subsection{Survival probabilities}
Let the initial excitation now be placed on the non-trap node $1$, i.e.,
$\langle 1 | \bm\rho_1(0) | 1\rangle=1$. From Eq.~(\ref{pavg}), the
survival probability follows as 
$\Pi(t)= \langle 1 | \bm\rho_1(t) | 1\rangle + \langle 2 |
\bm\rho_1(t) | 2\rangle $.
Without dephasing ($\lambda=0$) the survival probability is obtained by
direct diagonalization of $\bm H_S$, yielding the analytical result
\bea
\Pi(t) &=& \frac{e^{-\Gamma
t}}{|\textrm{cosh}(\phi_{\scriptscriptstyle\Delta})|^2}\Big[\left|\textrm{cosh}\left(t\cdot
f(\Gamma,\Delta,V)+\phi_{\scriptscriptstyle\Delta}\right)\right|^2
\nonumber \\
&&
+\left|\textrm{sinh}\left(t\cdot
f(\Gamma,\Delta,V)\right)\right|^2\Big],
\label{pil0}
\eea
with
$f(\Gamma, \Delta,
V)=\frac{1}{2}\left(\sqrt{(r-y)/2}+i\sqrt{(r+y)/2}\right)$,
where $r=\sqrt{y^2+4\Gamma^2\Delta^2}$ and $y=4V^2-\Gamma^2+\Delta^2$.
The limit $\Delta\to0$ yields 
\cite{mmsb2010}:
\be%a
\label{eq:nodeph_nodis}
\Pi(t) = \frac{e^{-\Gamma t}}{\cos^2
(\phi_{\scriptscriptstyle\Gamma})} \Big[\cos^2(Vt\cos(\phi_{\scriptscriptstyle\Gamma})-\phi_{\scriptscriptstyle\Gamma}) 
%\nonumber \\
%&& 
+ \sin^2(Vt\cos(\phi_{\scriptscriptstyle\Gamma}))\Big],
\ee%a
with $\phi_{\scriptscriptstyle\Gamma}=\textrm{arcsin}(\Gamma/2V)$. Letting
now also $\Gamma\to 0$, one arrives at the usual resonant Rabi oscillations for the
dimer \cite{Sakurai}.

Figure~\ref{surv-probs}(a) shows $\Pi(t)$ for $\lambda=0$ and for different values of
$\Delta$: Increasing $\Delta$ leads to a slower decay of $\Pi(t)$. While
small values $\Delta=0.1$ do not lead to strong deviations from the
$\Delta=0$ results, larger values of $\Delta$ clearly shift the whole
$\Pi(t)$ curve upwards. This can be interpreted as the rudimentary
onset of localization, the disorder being the off-set $\Delta$. Note also
that all curves have a plateau-like region where the slope decreases. By
increasing $\Delta$ this ``plateau'' is shifted to larger values of
$\Pi(t)$. This plateau is due to the fact that the dimer only has two
nodes. For larger networks, the averaging in $\Pi(t)$ diminishes such
effects.

%Coupling the system to the environment does not allow for an analytical solution anymore. However, in the limit of vanishing trapping strength ($\Gamma=0$) and for $\Delta\to0$ one has $\Pi(t)=1$ and obtains
%\bea \label{eq:notrap_nodis_1}
%\langle 1 | \bm\rho_1(t) | 1\rangle &=& \frac{1}{2}+\frac{e^{-\lambda t}}{2}\Big[\tan(\phi_{\scriptscriptstyle\lambda})\sin(2Vt\cos(\phi_{\scriptscriptstyle\lambda})) 
%\nonumber \\
%&& + \cos(2Vt\cos(\phi_{\scriptscriptstyle\lambda}))\Big]
%\eea
%with $\phi_{\scriptscriptstyle\lambda}=\textrm{arcsin}(\lambda/2V)$ and $\langle 2 | \bm\rho_1(t) | 2\rangle = 1 - \langle 1 | \bm\rho_1(t) | 1\rangle$.

Now, consider the case with a small trapping strength ($\Gamma\ll V$) and
let also the coupling be small ($\lambda\ll V$). By combining
the results for $\lambda=0$ and for $\Gamma=0$
%Eqs.~(\ref{eq:nodeph_nodis}) and (\ref{eq:notrap_nodis_1})
, and by expanding
all terms except exponentials to first order in $\Gamma$ and
$\lambda$, results in the simple exponential decay $\Pi(t)\approx
e^{-\Gamma t}$ for not too short times.

Plots of $\Pi(t)$ for different values of $\Delta\neq0$ and $\lambda\neq0$
are shown in Fig.~\ref{surv-probs}(b) and (c). The behavior is similar
both for small values $\Delta=0.1$ and larger values $\Delta=1.5$: An
increase of the coupling $\lambda$ leads always to a slower decay of
$\Pi(t)$. Moreover, for fixed $\lambda$, increasing $\Delta$ also leads to
a slower decay of $\Pi(t)$, compare, e.g., the solid curves in
Fig.~\ref{surv-probs}(b) and (c).
However, both for small and for larger $\Delta$, the curves do not stay
equidistant at all times.  Therefore, on different time scales the trapping efficiency,
here defined as the time it takes to decrease $\Pi(t)$ from one value to
another, can vary. 

\subsection{Lifetimes}
\begin{figure}[ht]
\centerline{\includegraphics[clip=,width=0.9\columnwidth]{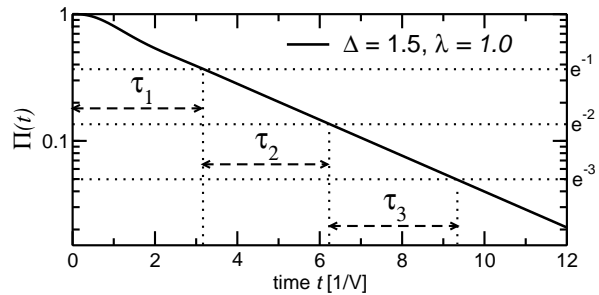}}
\caption{Definition of the lifetimes $\tau_1$, $\tau_2$,
and $\tau_3$, see text.}
\label{method}
\end{figure}

In order to obtain a quantitative measure for the efficiency of the
transport from node $1$ to the trap node $2$, we define different
lifetimes of the excitation, based on the decay of $\Pi(t)$ between two
given values:
\begin{itemize}
\item[$\tau_1$:]
decrease of $\Pi(t)$ from $1$ to $e^{-1}$,
\item[$\tau_2$:]
decrease of $\Pi(t)$ from $e^{-1}$ to $e^{-2}$,
\item[$\tau_3$:]
decrease of $\Pi(t)$ from $e^{-2}$ to $e^{-3}$.
\end{itemize}
see also Fig.~\ref{method}. Thus, after $t_1\equiv\tau_1$ about $63\%$ of the
probability has been absorbed by the trap, after $t_2\equiv\tau_1+\tau_2$ about
$86\%$, and after $t_3\equiv\tau_1+\tau_2+\tau_3$ about $95\%$. By considering the
lifetimes $\tau_s$ individually, we are able to quantify the rate
$1/\tau_s$ at which a certain amount of probability gets to be absorbed
given the initial value of $\Pi(t_{s-1})$.

Additionally, we also consider the time $\tau_\infty$, which is obtained
by rewriting the LQME, Eq.~(\ref{lvne}), in the form 
$\dot{\vec{\bm\rho}}_k(t) = {\cal L} \vec{\bm\rho}_k(t)$,
whose formal solution 
is ${\vec{\bm\rho}}_k(t) = \exp({\cal L}t)
{\vec{\bm\rho}}_k(0)$, where $\vec{\bm\rho}_k(t) = ( \langle 1 | \bm\rho_k(t) |
1\rangle,  \langle 1 | \bm\rho_k(t) |
2\rangle, \dots, \langle N | \bm\rho_k(t) |
N \rangle)$ is an $N^2$-dimensional vector.
Diagonalization of ${\cal L}$ then leads to
${\vec{\bm\rho}}_k(t) = {\cal Q} \exp(\bm \Lambda t) {\cal Q}^{-1}
{\vec{\bm\rho}}_k(0)$,
where ${\cal Q}$ is the eigenvector matrix of ${\cal L}$ and $\bm\Lambda$
a diagonal matrix whose elements are the eigenvalues of ${\cal L}$.
The long-time decay rate $1/\tau_\infty$ is then given by the smallest of
all the real parts of the eigenvalues of $\cal L$. We checked numerically
that in our case $\sum_n \int_0^\infty dt \langle n | \rho_k(t) | n\rangle$ is
identical to the results of Cao and Silbey for the mean first passage
time, see Eq.~(4) in \cite{cao2009}. We note, however, that here we focus
on the different lifetimes defined by the mean survival probability and
not on averaged quantities as the mean first passage time.

\begin{figure}[ht]
\centerline{\includegraphics[clip=,width=0.9\columnwidth]{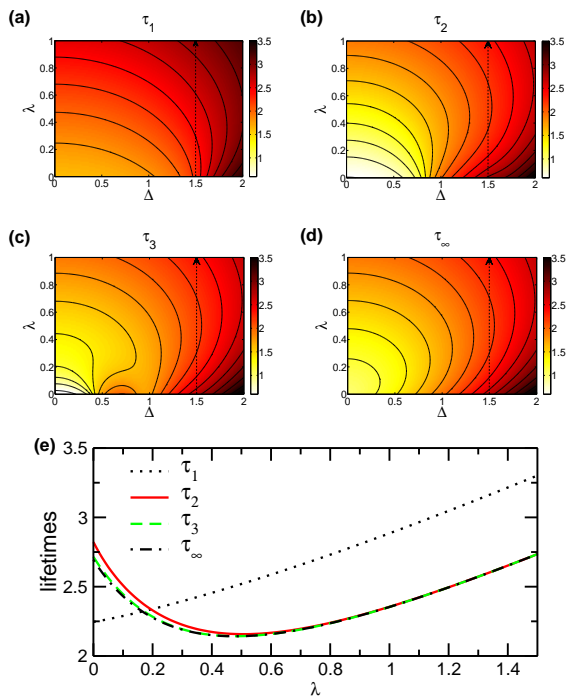}}
\caption{(Color online) (a)-(d) Contour plots of the lifetimes $\tau_1$, $\tau_2$,
$\tau_3$, and $\tau_\infty$, respectively, as functions of $\lambda$ and $\Delta$ for
$V=\Gamma=1$. The arrows at $\Delta=1.5$ show the location of the cuts for
fixed $\Delta$ and increasing $\lambda$ displayed panel (e).}
\label{lifetimes}
\end{figure}

Figures~\ref{lifetimes}(a)-(d) show contour plots of $\tau_1$, $\tau_2$,
$\tau_3$, and $\tau_\infty$, respectively, as functions of $\lambda$ and $\Delta$ for 
$V=\Gamma=1$. All figures show a similar behavior: 
\bi
\item[(i)]
For fixed
$\lambda$ and increasing $\Delta$ all lifetimes increase monotonically
(except $\tau_3$ for small $\lambda$, see below). 
\item[(ii)]
For not too small values of $\Delta$, however, all
lifetimes except $\tau_1$ (see below) first decrease with increasing
$\lambda$ to a minimum and then increase again.
\ei

The latter behavior is exemplified in
Fig.~\ref{lifetimes}(e) for $\Delta=1.5$. Clearly there is a minimum for
$\tau_2$, $\tau_3$, and $\tau_\infty$ at about the same value of
$\lambda\approx0.5$. Therefore, there is an optimal value of
$\lambda\neq0$ which leads to larger rates $1/\tau_s$ ($s\neq1$) compared
to the ones for $\lambda=0$. This implies, for instance, that for $\tau_2$
an additional $23\%$ of the probability gets to be absorbed by the trap
faster if the dimer is coupled to the environment. However, it does {\sl
not} neccessarily imply that $86\%$ of the probability transferred to the
trap after time $t_2$ gets to be absorbed faster. The same reasoning
holds also for $\tau_3$ and $\tau_\infty$. The initial absorption at the
rate $1/\tau_1$ cannot be enhanced for these values of $\Delta$ of the
order ${\cal O}(1)$ by the
coupling to the environment. 
This would mean that, in the interval $\tau_1$, (the later part of) a
$\Pi(t)$ curve for $\lambda'>\lambda$ would lie below the curve for
$\lambda$, which within our LQME approach is not the case, see
Figs.~\ref{surv-probs} (b) and (c). However, increasing $\Delta$ to larger
values also leads to a shallow minimum for $\tau_1$ as a function of
$\lambda$, see Fig.~\ref{lifetimes}(a) for $\Delta\approx 2$. 

The increase of the rates also depends on the value of $\Delta$. While an
increase of $\tau_\infty$ is visible even for small $\Delta\approx0.1$,
this is not the case for $\tau_2$ and $\tau_3$. Thus, in order to obtain a
sufficiently large increase of the rates $\tau_2$ and $\tau_3$, one needs
to have a value of $\Delta$ of the order ${\cal O}(1)$.
Intuitively, one observes a ``competition'' between localization and
decoherence effects: While the energy offset $\Delta$ tends to bind the
excitation at the initial node, the environmental coupling $\lambda$ tends
to spread the excitation evenly over the dimer. However, with increasing
values of $\lambda$ one approaches the quantum Zeno limit which also leads
to a slower decay of the survival probability and thus to larger
lifetimes.  Therefore, the minimal lifetimes can be viewed as being caused
by an optimal decoherence rate $\lambda$ destroying localization due to
$\Delta$. We note that our results are in agreement with the findings of
Cao and Silbey \cite{cao2009} and with the ones of Rebentrost {\sl et al.}
\cite{rebentrost2009}.

The fact that the lifetimes $\tau_s$ ($s\geq2$) lie on
the same curve for $\lambda\agt0.5$ translates to an (on average)
exponential decrease of the survival probability: 
Let the lifetimes $\tau_{s}$ be obtained piecewise, approximately via
$\Pi(t_{s-1}) - \Pi(t_{s}) \sim \exp(- a \tau_{s})$,
where $a$ is some constant. Now, if
$\tau_s= \tau$ for all $s$, we obtain
$\Pi(t_{s-r}) - \Pi(t_{s}) \sim \exp(- a r \tau)$,
with $r=1,\dots,s-1$. Allowing also $r=s$, one
has $\Pi(t_s) \sim \Pi(t_0) - \exp(-a t_s)$, where $t_s=s\tau$.

Finally, we remark on one peculiar feature of the behavior of $\tau_3$.
For small values of $\lambda\alt0.1$, the lifetime $\tau_3$ has a
local maximum at $\Delta\approx0.7$, visible as a small ``lagoon'' in
Fig.~\ref{lifetimes}(c). This is due to the plateau of $\Pi(t)$,
mentioned above. By increasing $\Delta$ for fixed (small) values of
$\lambda$ the plateau is shifted to larger values of $\Pi(t)$ and,
therefore, can eventually lie in the region between $\Pi(t)=e^{-1}$ and
$\Pi(t)=e^{-2}$. This then might lead to longer lifetimes $\tau_3$. The
other lifetimes $\tau_2$ and $\tau_\infty$ are not affected by this.

\section{Conclusions}
In conclusion, we have studied the trapping of excitations for a dimer with an
energy off-set and with environmental coupling. The survival probability,
$\Pi(t)$, not to get trapped (obtained from a Lindblad quantum master
equation) shows distinct features depending on the strength $\lambda$ of
the coupling to the environment and on the energy off-set $\Delta$: While
it is {\sl not} possible to always enhance the overall efficiency of
the decay to the trap, it is possible to increase the rate of trapping in
certain intervals of time, leading to faster decay of probability in these
intervals. The fastest decay is obtained for an optimal value of $\lambda$
which is about half the value of the coupling $V$ between the nodes.
However, a substantial increase is only obtained for values of $\Delta$ of
the order ${\cal O}(1)$.

%\begin{acknowledgments}
{\sl Acknowledgemetns. - }
Support from the Deutsche Forschungsgemeinschaft (DFG) and the Fonds der
Chemischen Industrie is gratefully acknowledged. 
We thank Alexander Blumen for continuous support and fruitful discussions.
%\end{acknowledgments}

%%%%%%%%%%%%%%%%%%%%%%%%%%%%%%%%%%%%%%%%%%%%%%%%%%%%%%%%%%%%%%%%%%%%%%%%%%%%%%%
% References
%%%%%%%%%%%%%%%%%%%%%%%%%%%%%%%%%%%%%%%%%%%%%%%%%%%%%%%%%%%%%%%%%%%%%%%%%%%%%%%
%\bibliographystyle{unsrt}

%\bibliography{/home/muelken/papers/bibliography/all}
\end{document}